\documentclass[useAMS,usenatbib]{mn2e} 
\usepackage{graphicx,epsfig}

\title[First results of the {\em XI} Groups Project]{First results of 
  the {\em XI} Groups Project: Studying an unbiased sample of galaxy 
  groups} \author[J. Rasmussen et al.]{Jesper 
  Rasmussen,$^{1}$\thanks{E-mail: jesper@star.sr.bham.ac.uk} Trevor J. 
  Ponman,$^{1}$ John S. Mulchaey,$^{2}$ Trevor A. Miles$^{1}$ 
  \newauthor 
  and Somak Raychaudhury$^{1}$\\ 
  $^{1}$School of Physics and Astronomy, University of Birmingham, 
  Edgbaston, 
  Birmingham B15 2TT\\ 
  $^{2}$Observatories of the Carnegie Institution, 813 Santa Barbara 
  Street, Pasadena, California, USA} 
\begin{document} 
 
\date{} 
 
\pagerange{\pageref{firstpage}--\pageref{lastpage}} \pubyear{2006} 
 
\maketitle 
 
\label{firstpage} 
 
\begin{abstract} 
  X-ray observations of hot, intergalactic gas in galaxy groups
  provide a useful means of characterizing the global properties of
  groups.  However, X-ray studies of large group samples have
  typically involved very shallow X-ray exposures or have been based
  on rather heterogeneous samples. Here we present the first results
  of the {\em XI} (XMM/IMACS) Groups Project, a study targeting, for
  the first time, a redshift-selected, statistically unbiased sample
  of galaxy groups using deep X-ray data.  Combining this with radio
  observations of cold gas and optical imaging and spectroscopy of the
  galaxy population, the project aims to advance the understanding of
  how the properties and dynamics of group galaxies relate to global
  group properties.  Here, X-ray and optical data of the first four
  galaxy groups observed as part of the project are presented.  In two
  of the groups we detect diffuse emission with a luminosity of
  $L_{\rm X} \approx 10^{41}$~erg~s$^{-1}$, among the lowest found for
  any X-ray detected group thus far, with a comparable upper limit for
  the other two.  Compared to typical X-ray selected groups of similar
  velocity dispersion, these four systems are all surprisingly X-ray
  faint.  We discuss possible explanations for the lack of significant
  X-ray emission in the groups, concluding that these systems are most
  likely collapsing for the first time.  Our results strongly suggest
  that, unlike our current optically selected sample, previous X-ray
  selected group samples represented a biased picture of the group
  population. This underlines the necessity of a study of this kind,
  if one is to reach an unbiased census of the properties of galaxy
  groups and the distribution of baryons in the Universe.
 
\end{abstract}

\begin{keywords} 
  galaxies: clusters: general -- galaxies: distances and redshifts -- 
  X-rays: galaxies -- X-rays: galaxies: clusters 
\end{keywords} 
 
\section{Introduction}\label{sec,intro} 
 
In hierarchical theories of structure formation, structures of 
progressively increasing size separate out from the Hubble expansion, 
recollapse and virialise. Groups of galaxies are the characteristic 
structures which have formed by the present epoch, and are believed to 
contain the bulk of the matter in the Universe \citep*{fuku98}. As 
such, an understanding of the Universe requires an understanding of 
galaxy groups. 
 
The depth of the potential wells of groups is similar to that of
individual galaxies, and galaxy velocities within groups are only a
few hundred km~s$^{-1}$. Under these circumstances, galaxies can
interact strongly with one another, and with the group potential
(e.g.\ \citealt{sers74,meno92,verd98,mend03,mile04}).  Hence, not only
are groups the most common environment for galaxies, but they provide
an environment which has a strong effect on galaxy properties, and
which is itself evolving, as groups collapse, virialise and grow
through mergers and accretion.
 
Despite the crucial role played by groups in cosmic structure
formation and evolution, they have received relatively little
attention compared to larger clusters. This is at least partly due to
the fact that typical groups may contain only a handful of bright
galaxies in their inner regions. In optical data, groups are therefore
often difficult to detect with confidence in projection.  X-ray
emission from a hot intragroup medium (IGM) provides a much more
reliable method of detecting the potential well of a virialised group.
Some earlier works based on pointed {\em ROSAT} X-ray observations
utilized this to study the global properties of groups and those of
their member galaxies in a coherent manner
\citep{mulc96,mulc98,zabl98,hels00,hels03,osmo04}.  However, although
care had been taken to include groups spanning a wide range of
properties, as in the study of \citet{osmo04}, from a statistical
viewpoint these works all suffered from the fact that {\em ROSAT} only
targeted a heterogeneous sample of hand-picked groups.
 
X-ray investigations of optically selected group samples include those
of \citet{mahd97}, and later \citet{mahd00}, who invoked data from the
{\em ROSAT} All-Sky Survey (RASS) to study a redshift-selected,
unbiased sample of groups in X-rays.  However, no detailed information
on the galaxy dynamics within these groups was extracted.
\citet{burn96} also employed RASS data to investigate the X-ray
properties of a sample of groups, drawn from the optically selected
and statistically complete poor cluster catalogue of \citet{ledl96} and
\cite{whit99}. This work furthermore included a detailed investigation
of the galaxy dynamics within the groups, though the systems were
originally selected by photometric enhancements and were not based on
velocity information.  RASS data also formed the basis for the search
for X-ray gas in the optically selected \citet{hick82} catalogue of
groups conducted by \citet*{ebel94}, a work later augmented by
\cite{ponm96}, who included pointed ROSAT data.  Even in the latter
study, however, the majority (roughly two-thirds) of the groups only
had RASS coverage.  Due to the shallow RASS exposures of typically a
few hundred seconds, the X-ray properties of groups with $L_{\rm X}
\la 10^{42}$~erg~$^{-1}$ could generally not be investigated in these
studies. Moreover, in many cases the data only allowed for a detection
of hot intragroup gas, whereas no detailed characterization of gas
properties could be carried out.  A further concern for RASS--based
results for the diffuse X-ray luminosity $L_{\rm X}$ from groups is
the potential danger of contamination from active galactic nuclei due
to the broad point spread function (PSF) of the {\em ROSAT} PSPC at
large off-axis angles.
 
With the exception of dedicated large-scale structure surveys (e.g.,
\citealt{pier04}), more recent observations with {\em XMM-Newton} and
{\em Chandra} have largely been targeting groups already studied by
{\em ROSAT} in pointed observations or serendipitously detected in
X-rays as part of the RASS. It is not clear to what extent these
groups, most of which are X-ray selected, can be seen as
representative of the overall group population at low redshift.
Studies of clusters of galaxies indicate that X-ray selection could
incorporate a serious bias, with X-ray and optically selected samples
showing highly disparate X-ray properties and in some cases
\citep{dona02} being largely non-overlapping.  \citet{gilb04} found
that a significant fraction of optically selected clusters do not have
a clear X-ray counterpart in {\em ROSAT} data, even though the
physical reality of the X-ray faint systems were confirmed by
spectroscopic follow-up. A similar conclusion, although based on
optical multi-colour imaging rather than spectroscopy, was reached by
\citet{bark06} using {\em Chandra} data.  The optically selected
high-redshift clusters of \citet*{lubi04} were found to be X-ray
underluminous for their velocity dispersion $\sigma_{\rm v}$,
deviating from the $L_{\rm X}$--$\sigma_{\rm v}$ relation derived for
rich, nearby X-ray clusters.  A similar result was shown for
colour-selected clusters by \citet{hick04}, while \citet{pope05} found
that a subset of their Abell clusters were X-ray underluminous with
respect to their virial mass. These results clearly suggest that X-ray
selection could provide a biased picture of the cluster population,
and there is little reason to assert that the situation for groups
should be significantly different. Given the lower X-ray luminosities
of groups, such a bias is likely to be even stronger for these
systems.
 
As a consequence of the fact that most X-ray studies of groups have
either been concentrating on X-ray selected targets picked from a
variety of different group catalogues, or have employed the shallow
RASS data to investigate group X-ray emission, we currently have no
{\em unbiased} census of the properties of hot gas and their relation
to the dynamics of galaxies within galaxy groups.  The advent of large
redshift surveys, coupled with the high sensitivity, spatial
resolution, and spectral capability of {\em XMM-Newton} and {\em
  Chandra}, has made it possible to remedy this situation. Groups can
be selected in redshift space, the properties of any hot IGM can be
assessed using deep X-ray data, and group dynamics can be studied with
multi-slit spectrographs.  We have launched a project to take
advantage of this situation, by targeting a statistically
representative sample of groups with both {\em XMM-Newton} and the
IMACS multi-object camera and spectrograph \citep{bige03}, installed
on the 6.5-m Baade/Magellan telescope at Las Campanas.
 
The primary aim of the {\em XI} Project is to understand the nature
and evolution of the galaxy group population, and to study the way in
which the properties of groups are related to those of their member
galaxies. By performing a sensitive test for the presence of any hot
IGM in the selected groups, one of the key outcomes of this project
will be a reliable estimate of the fraction of optically selected
groups which actually contain a hot IGM. Another issue to be addressed
is whether the properties of any hot X-ray gas can be used as an
evolutionary tracer of the dynamical state of a group.  The thrust of
the present paper is to provide an initial attack on these questions,
using the results obtained for the first four groups observed by {\em
  XMM} as part of the project. The large collecting area and field of
view of {\em XMM} makes it the ideal X-ray instrument for this study.
IMACS has a field of view perfectly matched to that of {\em XMM}
($\sim 30$~arcmin in both cases), and allows us to obtain several
hundred spectra over this entire field in a single exposure, down to
$M_B \approx -15$ at the selected sample redshift of $z\approx 0.06$.
 
We assume $\Omega_m=0.3$, $\Omega_{\Lambda}=0.7$, and a Hubble
constant of $H_0=70$~km~s$^{-1}$~Mpc. The virial radius $r_{\rm vir}$
of each group can then be identified with $r_{100}$, the radius
enclosing a mean density of 100 times the critical density $\rho_c$
\citep*{eke96}. If the dark matter density of the groups follows an
'NFW' profile \citep*{nava95} with concentration parameters in the
plausible range $c=5-20$ \citep{bull01}, we then have $r_{500}\approx
0.5r_{\rm vir}$.  In the adopted cosmology, the characteristic
redshift of our sample of $z=0.06$ corresponds to a luminosity
distance $D\approx 275$~Mpc, and 1~arcmin to $\approx 70$~kpc.

\section{Group sample selection}\label{sec,sample} 
 
A redshift-selected sample provides the best basis for a study of this
type.  X-ray selection is impractical, due to the lack of wide-area
high-sensitivity data, and would in any case fail to detect groups
which are collapsing but do not yet have virialised cores.  We
therefore selected our sample from the catalogue of 2209 groups
derived by \citet[hereafter MZ]{merc02} from a friends-of-friends
(FOF) clustering analysis (optimised by reference to cosmological
simulations) on the first ('100-K') data release from the 2dF Galaxy
Redshift Survey \citep{coll01}.
 
From this large group catalogue, we selected a sample of 25 systems
which satisfied the following additional criteria: (i) Redshift
$z=0.060-0.063$: This narrow slice puts all groups on an equal
observational footing at a redshift where the characteristic radius
$r_{200}$ of groups of $\sim 1$~Mpc corresponds to 13~arcmin and so is
well matched to both the {\em XMM} and IMACS fields of view. Using a
narrow redshift interval further has the advantage of circumventing a
typical problem associated with FOF-catalogues, namely that the FOF
linking length grows with $z$, implying that the properties of
FOF-selected groups may vary systematically with redshift.  (ii)
Velocity dispersion $\sigma_{\rm v} < 500$~km~s$^{-1}$: Limiting our
study to poor systems, which are the most common (containing more than
half of all galaxies, see \citealt{tull87} and \citealt{eke04}) and
provide the environment in which dynamical evolution is most rapid,
and the dispersion in observed properties is greatest.  (iii) Number
of spectroscopically confirmed 2dF member galaxies $N_{\rm gal}\ge 5$:
Reducing the danger of including groups which are not real physical
associations, by allowing only systems which incorporate at least five
2dF galaxies. $N$-body simulations show that FOF-groups with $N_{\rm
  gal}\ge 5$ almost always correspond to gravitationally bound
structures rather than being unbound density fluctuations (e.g.,
\citealt*{rame97}).  (iv) Redshift completeness: We avoid the edges of
the 2dF survey area, and regions with poor completeness in the 100-K
data release.
 
Within these constraints, we selected our sample of 25 groups entirely
at random. Since our aim is to explore the full population of
optically selected systems, we deliberately did not attempt to apply
any further constraints of regularity, optical luminosity or dynamical
status.  Hence our sample contains groups spanning a wide range of
properties, subject to having a number density contrast $\delta \rho/
\langle \rho \rangle \geq 80$ with respect to the mean galaxy number
density $\langle \rho \rangle$, which distinguishes them in the FOF
analysis of MZ. To illustrate this diversity in group properties,
Fig.~\ref{fig,sigmarv} shows the distribution of the full sample in
the ($r_{\rm vir},\sigma_{\rm v}$)--plane, with virial radii from
\citet{merc02} based on the projected distribution of galaxies.
Highlighted are the four groups targeted for the initial study
described in this paper. Basic optical properties of these four groups
are provided in Table~\ref{tab,opt}.  Note that the subsample studied
here includes the two groups with the largest values of $\sigma_{\rm
  v}$ in the full {\em XI} sample, and that, within the constraints
imposed by our selection criteria, the {\em XI} sample covers most of
the available parameter space.  Also note that, despite the fact that
some of the groups are rather compact as judged from the MZ virial
radius, we find that none of the {\em XI} groups would satisfy the
compact group criteria introduced by \citet{hick82}. This is in all
cases because they fail to meet the compactness criterion
$\mu_R<26.0$, where $\mu_R$ is the total $R$-band magnitude per
arcsec$^{2}$ of the galaxies within 3~mag of the brightest one,
averaged over the (circular) angular extent of the subgroup defined by
these galaxies.

\begin{figure} 
\mbox{\hspace{-6mm} 
 \includegraphics[width=93mm]{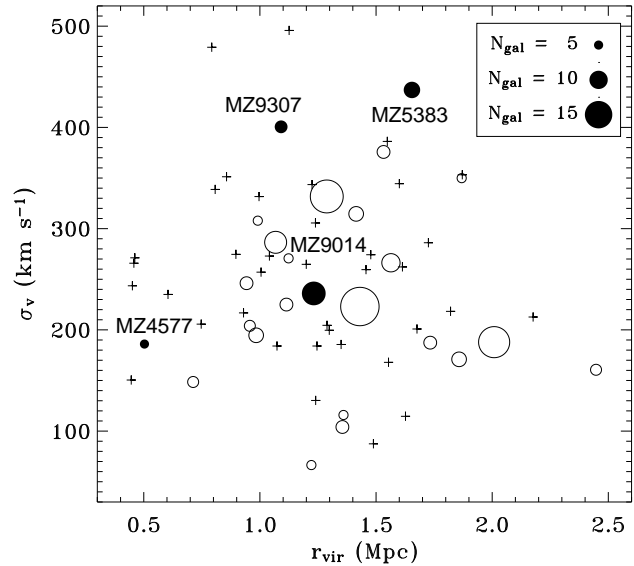}} 
\caption{Virial radii and velocity dispersions of all groups
  satisfying our selection criteria. Groups studied in this paper are
  marked by filled circles, with the remaining systems in the {\em XI}
  sample shown as empty circles.  For the {\em XI} groups, the size of
  each symbol is proportional to the number of member galaxies in each
  group, as derived by MZ.  Velocities have been updated where new
  IMACS values are available. Remaining groups, meeting our selection
  criteria but not included in the {\em XI} sample, are marked by
  small crosses.}
\label{fig,sigmarv} 
\end{figure}

\begin{table*} 
 \centering 
 \begin{minipage}{125mm} 
   \caption{Basic group properties, including number of confirmed 
     member galaxies $N_{\rm gal}$, mean recession velocities $\langle 
     v\rangle$, velocity dispersions $\sigma_{\rm v}$, and virial 
     radii $r_{\rm vir}$ from MZ. (*) = updated values from IMACS.} 
  \label{tab,opt} 
  \begin{tabular}{@{}lcccccc@{}} \hline \multicolumn{1}{l}{Group} & 
    \multicolumn{1}{c}{RA} & \multicolumn{1}{c}{Dec} & 
    \multicolumn{1}{c}{$N_{\rm gal}$} & \multicolumn{1}{c}{$\langle 
      v\rangle$} & \multicolumn{1}{c}{$\sigma_{\rm v}$} & 
    \multicolumn{1}{c}{$r_{\rm vir}$} \\ 
    & (J2000) & (J2000) & & (km~s$^{-1}$) & (km~s$^{-1}$) & (Mpc) \\ \hline 
    MZ~4577  & 11 32 30.79 & $-04$ 00 00.8 & 13 (*) & 18614 (*)  
             & $186^{+136}_{-28}$ (*) & $0.50\approx 7$~arcmin \\ 
    MZ~5383  & 12 34 52.83 & $-03$ 35 54.3 & 23 (*) & 18120 (*)  
             & $437^{+92}_{-43}$ (*) & $1.65\approx 23$~arcmin \\ 
    MZ~9014  & 00 37 48.12 & $-27$ 30 29.1 & 22 (*) & 18239 (*)  
             & $236^{+80}_{-45}$ (*) & $1.23\approx 17$~arcmin \\ 
    MZ~9307  & 00 40 48.64 & $-27$ 27 06.1 &  7 & 18252 & 401  
             & $1.09\approx 15$~arcmin \\ 
    \hline 
\end{tabular} 
\end{minipage} 
\end{table*}

Since we initiated the {\em XI} project, the 2dF Galaxy Redshift
Survey has been completed, and a clustering analysis similar to that
of MZ has been carried out on the full catalogue\footnote{Available
  at http://www.mso.anu.edu.au/2dFGRS/}, resulting in the 2PIGG group
catalogue \citep{eke04}. For groups with $N_{\rm gal}\ge 4$,
\citet{eke04} find excellent agreement with MZ regarding
catalogue-averaged group redshift and velocity dispersion, and the
fraction of galaxies grouped. To test for the presence of significant
differences in properties between groups specifically fulfilling our
selection criteria in the two catalogues, we selected from both of
these all groups within the richness range $N_{\rm gal} = 5-24$
occupied by the {\em XI} groups in the MZ catalogue, and a
three-dimensional velocity dispersion $\sigma <500$ km/s. To obtain
sufficient statistics for a useful comparison, we adopted a wider
redshift range $0.04 < z < 0.08$ on either side of the {\em XI} group
redshifts.

The histograms plotted in Fig.~\ref{fig,2dfgrs} show the resulting
distributions of number of member galaxies and velocity dispersion of
the 899 2PIGG groups (solid line) and 402 MZ groups (dotted line) that
were found in this parameter range. The ratio of the number of groups
extracted from the two catalogues, 2.24, is very similar to that of
the total number of galaxies in the catalogues. For ease of
comparison, the histograms also show the MZ sample scaled by 2.24.
The resulting richness histograms for the 2PIGG and (scaled) MZ
samples are seen to be almost identical. From a Kolmogorov--Smirnov
(K--S) test, the probability that these two samples are significantly
different is $P=3\times 10^{-7}$.  The velocity dispersion histograms
peak at roughly the same location, though there is some detailed
difference in the structure of the peaks, possibly related to the
difference in methods of calculating $\sigma_{\rm V}$ between the two
catalogues. From a K--S test, the probability that these two samples
are significantly different is only $P=0.044$ however, so we conclude
that the 2PIGG and MZ subsamples satisfying our selection criteria are
not significantly different.

\begin{figure} 
\mbox{\hspace{-0mm} 
\rotatebox{-90}{ 
 \includegraphics[height=83mm]{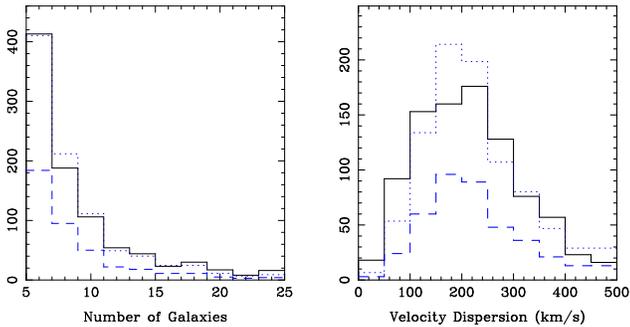}}} 
\caption{Histograms showing the number of groups versus $N_{\rm gal}$
  (left) and $\sigma_{\rm V}$ (right) for groups fulfilling our
  selection criteria in the full 2PIGG group catalogue (solid lines)
  and the MZ catalogue (dashed lines). The dotted lines represent the
  MZ subsample scaled to the number of groups in the 2PIGG subsample.}
\label{fig,2dfgrs} 
\end{figure}

\section{Observations and analysis}\label{sec,obs}

\subsection{Optical data}

To obtain photometry, colours, and morphologies of galaxies in the
group fields, and to identify candidates for spectroscopic follow-up,
images of all groups in the Bessel {\em BVR} filters were obtained
with the $2048\times2048$~pixel Wide Field Reimaging CCD (WFCCD) on
the 100-inch du~Pont telescope at Las~Campanas. The field of view of
25~arcmin gives a plate scale of 0.77~arcsec~pixel$^{-1}$. Twelve
dithered exposures were taken for each group and filter, with typical
integration times for a group totalling 24~min in $V$ and $R$ and
60~min in $B$. Images of each group were median-combined and processed
using standard {\sc iraf} packages, with domeflats used to flatfield
the images.  Objects were identified using the {\sc SExtractor}
package \citep{bert96}, and detections were checked visually. Objects
with {\sc SExtractor}'s stellarity index $>0.9$ were deemed to be
definitely stellar and therefore not subject to further analysis. All
objects with full-width at half maximum less than the PSF were
discarded as noise.  A fixed aperture, set to be slightly greater than
the seeing, was used to obtain magnitudes in all filters. Objects in
different filters were matched, and aperture magnitudes were
subtracted to derive colours.
 
Multi-object spectroscopy of galaxies in the group fields was
performed with the IMACS spectrograph on the Baade/Magellan telescope,
using short (f/2) camera mode with a grism of 300 lines~mm$^{-1}$,
giving a wavelength range of 3900--10000~\AA~and a dispersion of
1.34~\AA~pixel$^{-1}$.  Typical exposure times were 2~hours for each
slit mask.  Spectroscopic candidates were selected from galaxy lists
generated from the $R$-band images taken at the du Pont telescope.
Priority was given to the brightest objects in each field, and colour
information was not used to select objects.  The IMACS data were
reduced using a set of programs developed by A.~Oemler. First,
overscan regions of the CCD chips were used to estimate and subtract
the bias level from each frame. Domeflat exposures were then used to
flatfield the data. Sky subtraction was performed using the procedure
described in \citet{kels03}.  Finally, wavelength calibrations were
determined for each spectrum from exposures of a He-Ne-Ar lamp.
 
At present we have spectroscopic coverage of 17 of the 25 groups,
including three of the four systems discussed here, MZ~4577, 5383 and
9014.  Galaxy velocities in these groups were measured by
cross-correlating the spectra with galaxy templates as described in
\citet{zabl98}.  Typical errors from the template fitting were $\delta
v\approx 50$~km~s$^{-1}$.  From a list of objects with velocities
within 2000~km~s$^{-1}$ of the group redshift, we determined group
membership and velocity dispersion $\sigma_{\rm v}$ by using the
bi-weight estimator of $\sigma_{\rm v}$ \citep*{beer90} to iteratively
discard $3\sigma$ outliers from the group.  The $1\sigma$ errors on
the resulting velocity dispersion were estimated from 10,000 bootstrap
trials, as described in \citet{beer90}.

\subsection{X-ray data}\label{sec,X-ray} 
 
The four groups discussed in this paper were all observed by {\em
  XMM-Newton} for the nominal exposure time of $\sim 20$~ks chosen for
the {\em XI} Project. This value is mainly driven by the need to
robustly test for the existence of a hot IGM even in groups with X-ray
luminosities among the lowest known, $L_{\rm X} \la
10^{41}$~erg~s$^{-1}$.  The {\em XMM} observation log is presented in
Table~\ref{tab,X} which details the observing modes and cleaned
exposure times for each EPIC camera.

\begin{table} 
 \centering 
   \caption{Summary of {\em XMM-Newton} observations. Column~3 specifies the 
frame mode (full frame/extended full frame) and optical blocking filter.} 
  \label{tab,X} 
  \begin{tabular}{@{}lcccc@{}} \hline 
    \multicolumn{1}{l}{Group} & 
    \multicolumn{1}{c}{EPIC} & 
    \multicolumn{1}{c}{Obs.\ mode} & 
    \multicolumn{1}{c}{$t_{\rm exp}$} & 
    \multicolumn{1}{c}{$N_{\rm H}$} \\  
    & & & (ks) & ($10^{20}$~cm$^{-2}$) \\ \hline 
    MZ~4577 & pn   & FF-Thin    &  2.2 & 4.08   \\ 
    \ldots  & MOS1 & FF-Thin    &  3.8 & \ldots \\ 
    \ldots  & MOS2 & FF-Thin    &  3.9 & \ldots \\ 
    MZ~5383 & pn   & FF-Thin    &  7.5 & 2.41   \\ 
    \ldots  & MOS1 & FF-Thin    & 10.3 & \ldots \\ 
    \ldots  & MOS2 & FF-Thin    & 10.7 & \ldots \\ 
    MZ~9014 & pn   & FF-Thin    & 22.3 & 1.54   \\ 
    \ldots  & MOS1 & FF-Thin    & 26.4 & \ldots \\ 
    \ldots  & MOS2 & FF-Thin    & 26.6 & \ldots \\ 
    MZ~9307 & pn   & EFF-Medium &  2.7 & 1.48   \\ 
    \ldots  & MOS1 & FF-Medium  &  7.7 & \ldots \\ 
    \ldots  & MOS2 & FF-Medium  &  8.9 & \ldots \\ 
    \hline 
\end{tabular} 
\end{table}

The {\em XMM} data were analysed using {\sc xmmsas} v6.0, and
calibrated event lists were generated with the {\sc emchain} and {\sc
  epchain} tasks.  Event files were filtered using standard quality
flags, while retaining only patterns $\leq 4$ for pn and $\leq 12$ for
MOS.  Screening for background flares was first performed in the
10--15~keV band for MOS and 12--14 keV for pn. Following an initial
removal of obvious large flares, a 3$\sigma$ clipping of the resulting
lightcurve was applied.  Point sources were then identified by
combining the results of a sliding-cell search ({\sc eboxdetect}) and
a maximum likelihood point spread function fitting ({\sc emldetect}),
both performed in five separate energy bands to span the range 0.3--12
keV.  In order to filter out any remaining soft protons in the data, a
second lightcurve ($3\sigma$) cleaning was then done in the
0.4--10~keV~band, within a 9--12~arcmin annulus which excluded the
detected point sources.  Closed-filter data from the calibration
database and blank-sky background data \citep{read03} for the
appropriate observing mode were filtered similarly to source data, and
screened so as to contain only periods with count rates within
$1\sigma$ from the mean of the source data.  All point sources were
excised out to at least 25~arcsec in spectral analysis.
 
To aid the search for diffuse X-ray emission within the groups,
smoothed exposure-corrected images were produced, with background maps
generated from blank-sky data. We allowed for a differing contribution
from the non-vignetted particle background component in source- and
blank-sky data by adopting the following approach.  First an EPIC
mosaic image was smoothed adaptively ($3\sigma$--$5\sigma$
significance range), and the particle background was subtracted. The
latter was estimated from closed-filter data which were scaled to
match source data count rates in regions outside the field of view,
and smoothed at the same spatial scales as the source data. The
resulting photon image includes the X-ray background at the source
position. To remove this component, a particle-subtracted blank-sky
image was produced in a similar way, and scaled to match 0.3--2~keV
source count rates in a point-source--excised 10--12~arcmin annulus
assumed to be free of IGM emission.  This image was then subtracted
from the corresponding source image, and the result was finally
exposure-corrected using a similarly smoothed exposure map.
 
Resulting 0.3--2~keV images of the region of each group covered by all
three EPIC cameras are shown in Fig.~\ref{fig,smoothed}, along with
the position of the optical group centre. Rather than adopting the
coordinates derived by MZ for the latter (listed in
Table~\ref{tab,opt}), these being just a straight mean of the galaxy
positions, we have indicated the luminosity-weighted centre instead,
calculated from the updated member lists where relevant.  It is
immediately apparent from these images that none of the groups
resembles the relaxed, X-ray bright systems characteristic of X-ray
selected group samples.
\begin{figure*} 
\begin{center} 
 \mbox{ 
 \includegraphics[width=140mm]{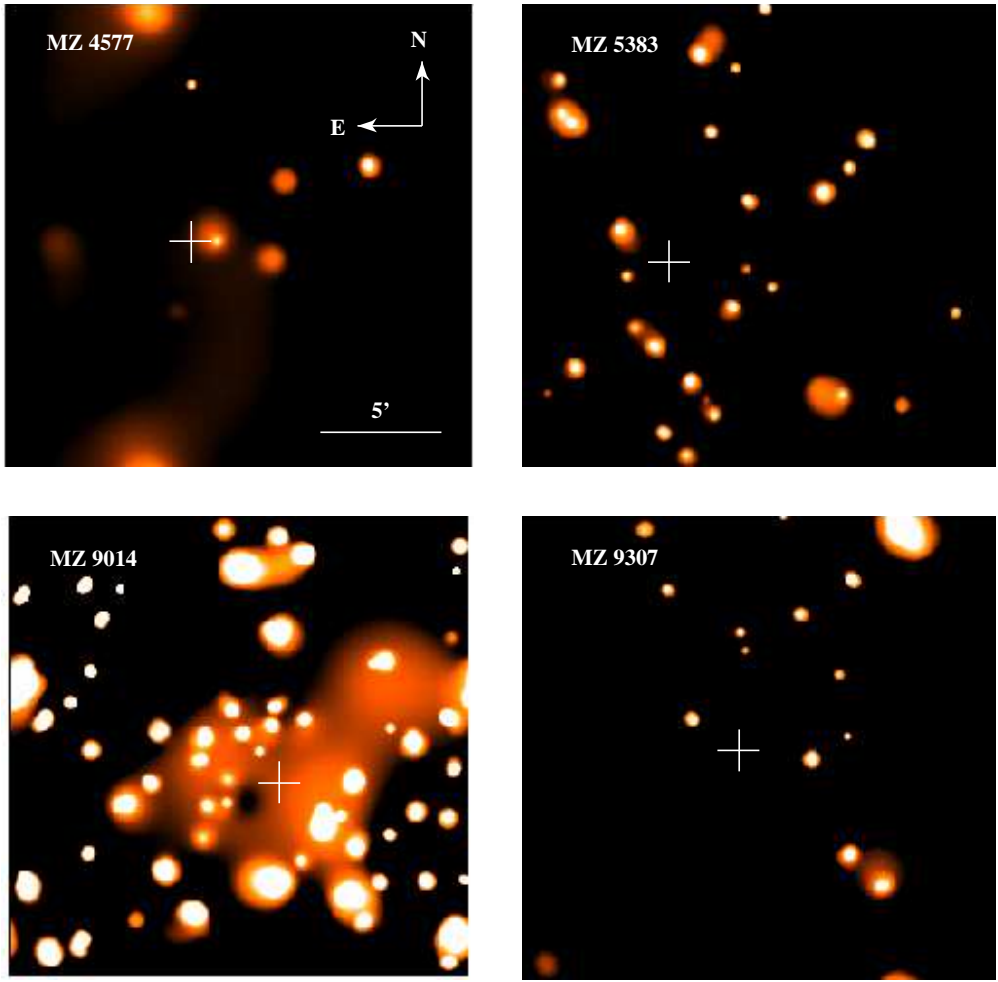}} 
 \caption{Adaptively smoothed 0.3--2~keV EPIC mosaic images of the central  
$19\times 19$~arcmin$^2$ of each group. A cross marks the position of 
the luminosity-weighted optical group centre in each case.} 
\label{fig,smoothed} 
\end{center} 
\end{figure*} 
 
Using smoothed and unsmoothed versions of the X-ray maps of
Fig.~\ref{fig,smoothed} along with the significance maps from the
smoothing procedure, we searched for evidence of diffuse emission on
at least two scales.  As earlier {\em ROSAT} studies showed that most
of the detectable IGM emission in X-ray bright groups is typically
concentrated within $\sim 200$~kpc \citep{hels00}, a natural first
step was to look for emission within this region, centred on the
optical group centre.  However, groups in the early stages of collapse
will not yet have virialised cores but could still contain
shock-heated X-ray gas distributed on larger scales, so we also
searched for hot gas within $0.5r_{\rm vir}$ (where significantly
different from 200 kpc), with virial radii taken from \citet{merc02}.
As mentioned in Section~\ref{sec,intro}, this radius is expected to
correspond to $r_{500}$ for a virialised group, which is the maximum
radius out to which group emission has so far been reliably detected
with {\em XMM} \citep{rasm04}. Characterizing the X-ray properties
inside this region also allows for a straightforward comparison to the
results of \citet{osmo04}, who derived X-ray luminosities inside
$r_{500}$ for their sample of 60 groups, by extrapolating fitted
surface brightness profiles out to this radius.
 
We note that the 10--12~arcmin annulus used to evaluate the local soft
X-ray background in the imaging data could in principle contain some
IGM emission, despite covering a region beginning at a radius
comparable to, or greater than, the adopted value of $r_{500}$ for all
four groups. If so, some IGM emission would erroneously be subtracted
from the images along with the local X-ray background. However, large
amounts of IGM emission beyond $r_{500}$ in these seemingly unrelaxed
systems would be a real surprise.  Indeed, as will be discussed, we do
not detect diffuse emission beyond 10~arcmin from the optical group
centre for any of the groups, so a 10--12~arcmin annulus is free of
detectable IGM emission in all cases (note that
Fig.~\ref{fig,smoothed} shows only the region covered completely by
all three EPIC cameras, and that the adopted annulus is largely
outside this region).  To help verify this, a 12--14~arcmin annulus
was used for comparison, at the expense of losing some detector area
within the field of view, resulting in poorer background statistics.
Within the Poisson errors, this did not change the number of detected
IGM photons for any of the groups.

For the spectral analysis of extended emission, the background was
evaluated by means of the common 'double-subtraction' technique
\citep{arna02}, using blank-sky background data from \citet{read03}
for the on-chip background, and a large-radius (10--12~arcmin) annulus
for determining the local soft X-ray background. For point sources,
surrounding 0.5--1~arcmin annuli in the source data were used for
background estimates.  X-ray spectra were accumulated into bins of at
least 20 net counts, and fitted in {\sc xspec} v11.0. Where
appropriate, thermal ({\sc mekal}) plasma model fits were used to
estimate X-ray luminosities $L_{\rm X}$ and the mean cooling time
$\langle t_{\rm cool}\rangle \approx 3kT/(\Lambda \langle n_e
\rangle)$ of X-ray gas associated with the X-ray detected galaxies and
with the IGM itself. Here, $\langle n_e \rangle \sim (EM/V)^{1/2}$ is
the mean electron density as inferred from the fitted emission measure
$EM$ and the assumed volume $V$. We have adopted the cooling function
$\Lambda(T,Z)$ of \citet{suth93}.  Where a reliable IGM temperature
measurement could not be obtained, $T_{\rm X}$ was taken from the
$\sigma_{\rm V}$--$T_{\rm X}$ relation of \citet{osmo04},
\begin{equation} 
  \mbox{log}\, \sigma_{\rm V} = (1.15\pm0.26)\mbox{ log}\, T_{\rm X} +  
  2.60\pm 0.03,
\label{eq,sigma_T} 
\end{equation} 
with $\sigma_{\rm V}$ in km~s$^{-1}$ and $T_{\rm X}$ in keV. Errors on
$T$ were derived from the dispersion of this relation, adding in
quadrature the error on $\sigma_{\rm V}$ itself. The latter was taken
from Table~\ref{tab,opt} where available, while a 20~per~cent error
was assumed for MZ~9307. The resulting temperature range, in
combination with the conservative assumption of a plasma metallicity
anywhere in the range 0--1~Z$_\odot$, was then used to estimate limits
to $L_{\rm X}$, $\langle n_e \rangle$, $\langle t_{\rm cool}\rangle$,
and total hot gas mass $M_{\rm gas}$.  For X-ray undetected galaxies,
only flux and luminosity limits were computed, in all cases assuming a
power-law spectrum of photon index $\Gamma=2$.

\section{Results} 
 
Optical $R$-band images of the groups are presented in
Fig.~\ref{fig,X-opt}, along with X-ray contours from
Fig.~\ref{fig,smoothed}.  Radial X-ray surface brightness profiles,
extracted from the particle-subtracted data with point sources masked
out, are shown in Fig.~\ref{fig,surfbright}. With the aid of these
figures, the X-ray and optical results for each group are discussed in
detail below.

\begin{figure*} 
\begin{center} 
 \mbox{ 
 \includegraphics[width=150mm]{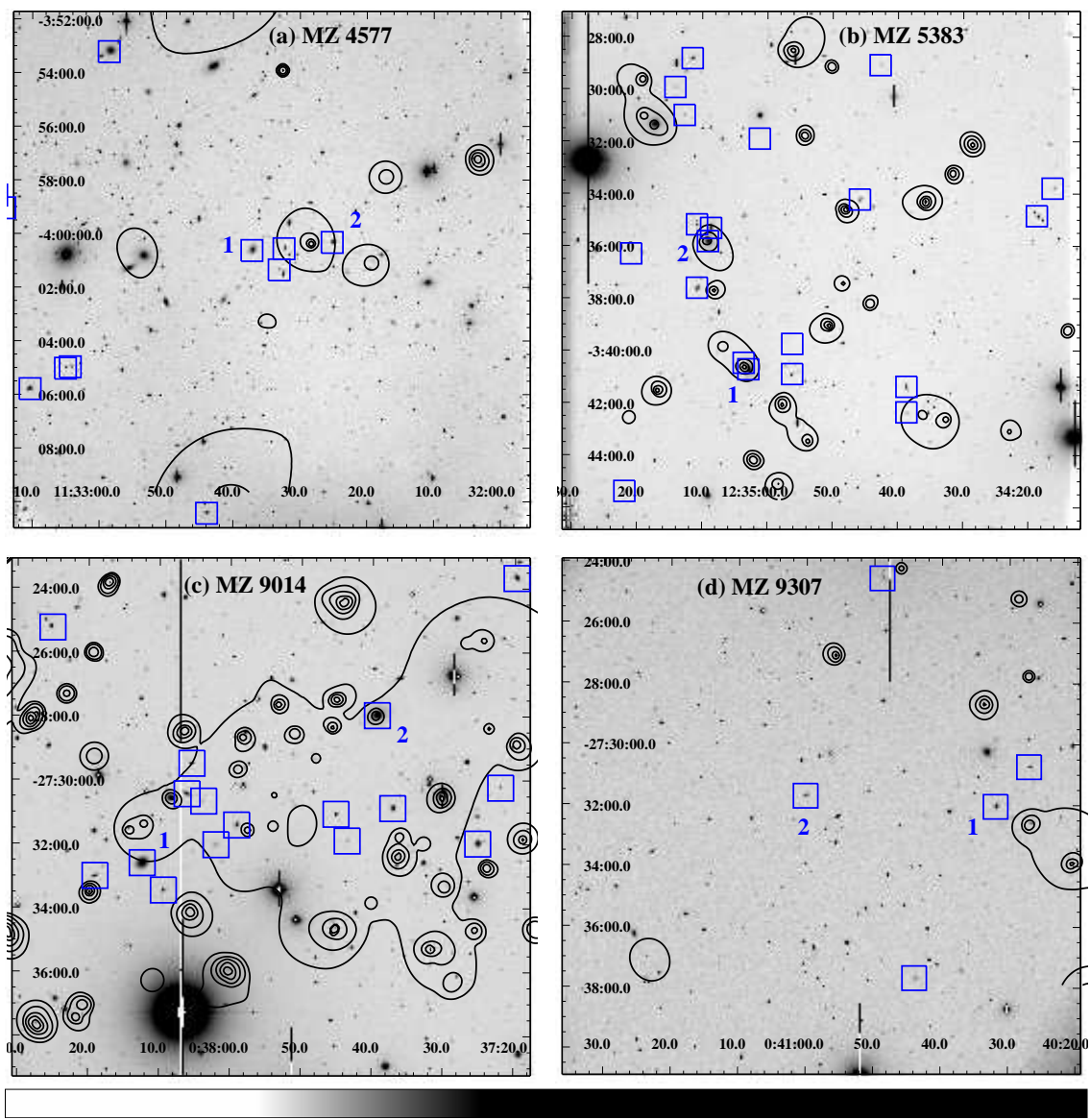}} 
\caption{X-ray contours of Fig.~\ref{fig,smoothed} overlayed on
  du~Pont $R$-band images of (a) MZ~4577, (b) MZ~5383, (c) MZ~9014,
  and (d) MZ~9307.  X-ray contours are logarithmically spaced over two
  decades, beginning at (a) 0.5, (b) 1.0, (c) 0.2, and (d) $1.8 \times
  10^{-6}$~photons~s$^{-1}$~arcsec$^{-2}$, respectively.  Confirmed
  galaxy members are marked by squares, with the first and
  second-ranked galaxy labelled. For MZ~4577 and MZ~9014, the
  outermost contour roughly corresponds to the regions inside which
  X-ray emission is detected at $\geq 3\sigma$ significance.}
\label{fig,X-opt} 
\end{center} 
\end{figure*}

\subsection{MZ~4577} 
 
From the analysis of \citet{merc02}, the virial radius of this group
is only $r_{\rm vir}\approx 500$~kpc. This makes it the most compact
group in our sample of 25 systems, enabling a search for diffuse X-ray
emission inside the full virial radius in our {\em XMM} observation.
As will be discussed in Section~\ref{sec,discuss}, the velocity
distribution of the 13 spectroscopically identified group galaxies
shows a bimodal structure, a fact which underlies the large upper
error on our derived value of $\sigma_{\rm V}$ listed in
Table~\ref{tab,opt}.  The updated velocity dispersion from our IMACS
data is none the less consistent with the value $\sigma_{\rm
  V}=223$~km~s$^{-1}$ listed by MZ on the basis of just five galaxies.

Although a significant fraction of the X-ray exposure was affected by
background flares, useful constraints on both diffuse and point-like
X-ray emission in the group could still be obtained.  We detect seven
point sources within the {\em XMM} field, down to a limiting 0.3--2
keV flux of $\sim 3\times 10^{-15}$~erg~cm$^{-2}$~s$^{-1}$. None of
the confirmed group galaxies is detected in X-rays, implying a
$3\sigma$ upper limit to their X-ray luminosities of $L_{\rm X} \la
2.7\times 10^{40}$~erg~s$^{-1}$ inside a 1~arcmin diameter circle.
Regarding diffuse X-ray emission in this group,
Fig.~\ref{fig,smoothed} does not reveal evidence for any clear
enhancement in X-ray surface brightness within $r_{\rm vir}\simeq
7$~arcmin, and the background-subtracted emission level inside $r_{\rm
  vir}$ is indeed found to be consistent at $1\sigma$ with the level
outside this region.  Nevertheless, there is suggestive evidence for
diffuse emission closer to the optical group centre.  This is
confirmed by the surface brightness profile shown in
Fig.~\ref{fig,surfbright}, centred on the peak of diffuse emission at
($11^{\rm h} 32^{\rm m} 29\fs 10$, $-04\degr 00\arcmin 11\farcs 4$),
which is roughly 50~arcsec east of the luminosity-weighted optical
group centre.  The profile demonstrates the presence of low-level
excess emission inside $\sim 50$~arcsec from the group centre (note
that all X-ray point sources, including the one seen close to the
optical group centre in Figs.~\ref{fig,smoothed} and \ref{fig,X-opt},
have been masked out when constructing this profile).  Quantitatively,
a comparison of the source and background X-ray maps described in
Section~\ref{sec,X-ray}, indicates the presence of diffuse emission
inside $0.5r_{\rm vir}$ at $2.4\sigma$ significance, relative to the
background level derived either immediately outside (250--400~kpc) or
within a large-radius annulus outside $r_{\rm vir}$.  The smoothing
significance map also confirms the presence of marginally significant
emission within this central region.
 
At a level of only $\sim 40$~net counts, the diffuse emission is too
faint to allow any useful spectral analysis. When combined with our
errors on $\sigma_{\rm v}$ from Table~\ref{tab,opt}, the $\sigma_{\rm
  v}$--$T_{\rm X}$ relation for X-ray bright groups,
equation~(\ref{eq,sigma_T}), would suggest $T =0.5^{+0.4}_{-0.1}$~keV
for the temperature of the IGM in this group. This temperature range
would translate into an unabsorbed 0.3--2~keV X-ray luminosity inside
$0.5r_{\rm vir}$ of $2.1 \pm 1.0\times 10^{41}$~erg~s$^{-1}$ at
90~per~cent confidence for any subsolar metal abundance $Z\leq
$~Z$_\odot$.  The derived luminosity can be compared to the value of
$L_{\rm X} \approx 9 \times 10^{41}$~erg~s$^{-1}$ inside $r_{500}$
suggested by the $L_{\rm X}$--$\sigma_{\rm v}$ relation of
\citet{osmo04}.  Thus, the group is X-ray underluminous relative to
the expectation from more X-ray bright systems. Under the above
assumptions on $T$ and $Z$, the flux measured inside $0.5r_{\rm vir}$
would formally imply a mean hot gas density within the range $\langle
n_e \rangle = 0.7-1.6\times 10^{-4}$~cm$^{-3}$, a gas mass
$1.3-3.1\times 10^{11}$~M$_\odot$ (i.e.\ well below the
$10^{12}$--$10^{13}$~M$_\odot$ typical for X-ray selected groups), and
a cooling time in the range 2.4--$9.7\times 10^{10}$~yr.  In
Table~\ref{tab,results}, we list some of these results, along with the
corresponding constraints obtained for the optically brightest galaxy
in the group.
 
We note that the diffuse emission is unlikely to represent an
undetected point source, given its spatial extent and the fact that it
is four times brighter than the faintest detected point source in the
field. Having no obvious optical counterpart, its centre is displaced
from any of the 2dF galaxies by more than 55~arcsec ($\sim 65$~kpc),
and so is also unlikely to represent emission from a galactic halo
associated with an elliptical within the 2dF redshift range of $z\leq
0.25$.  A further argument against association with an elliptical
comes from the typical ratio of $L_{\rm X}/L_B$ of ellipticals
\citep*{osul03}, which would suggest $L_B \approx 6\times
10^{10}$~L$_\odot$, corresponding to $m_B\approx 15.5$.  A galaxy of
this optical magnitude would be easily visible in
Fig.~\ref{fig,X-opt}a, where the brightest 2dF galaxy has $m_B=16.77$.
 
It is finally worth noting that the luminosity-weighted group centre
of MZ~4577 is only 5~arcmin (projected distance $\sim 350$~kpc) from
the centre of the Abell cluster A1308, situated at a redshift of
$z=0.0506$. MZ~4577 could currently be in the process of falling into
A1308. However, the radial velocity difference $\Delta v_r$ between
the two structures is roughly 3,600~km~s$^{-1}$, implying $\Delta v_r
\ga 4.9 \sigma_{\rm cl}$ for the measured cluster velocity dispersion
of $\sigma_{\rm cl}=652\pm 90$~km~s$^{-1}$ \citep{depr02}.  Even if
assuming the systems to be at exactly the same distance and having
vanishing transverse peculiar motions, the kinetic energy of the
combined system would exceed the gravitational binding energy by at
least a factor of 5 (for an assumed cluster mass of
$10^{14}$~M$_\odot$). It therefore seems unlikely that MZ~4577 is
falling into A1308.

\begin{figure} 
\begin{center} 
\mbox{\hspace{-2.5mm} 
 \includegraphics[width=89mm]{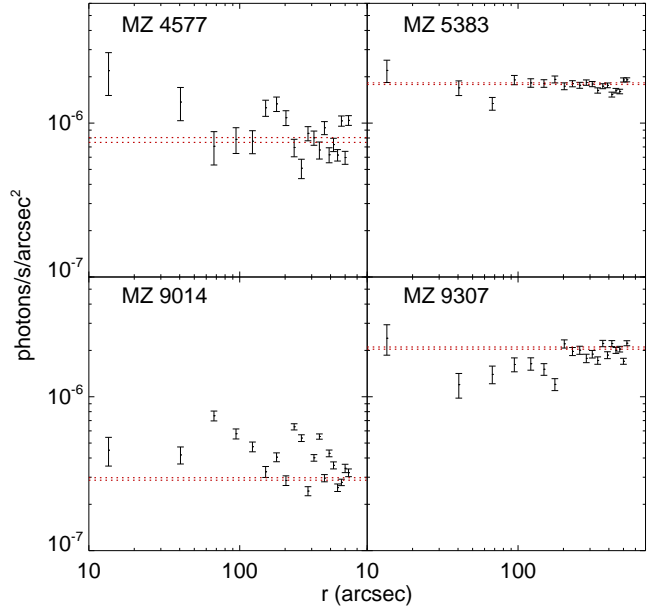}} 
\caption{0.3--2~keV exposure-corrected surface brightness profiles of
  the diffuse emission in all groups, extracted in 20 equal-size
  radial bins.  Dotted lines show the $1\sigma$ errors on the
  background level as estimated from a surrounding 9--11~arcmin
  annulus.}
\label{fig,surfbright} 
\end{center} 
\end{figure}

\subsection{MZ~5383} 
 
The distribution of galaxies within this group indicates a relatively
large virial radius of 23~arcmin from the MZ~analysis. The velocity
dispersion is also the largest in the full {\em XI} sample.  Our IMACS
results show that one of the 23 spectroscopically confirmed member
galaxies is offset from the mean velocity by roughly
$1100$~km~s$^{-1}$ (see Section~\ref{sec,discuss}), but eliminating it
from the sample would still leave $\sigma_{\rm v}$ consistent with the
value listed in Table~\ref{tab,opt}. It is not included in the 2dF
catalogue and therefore does not contribute to the value $\sigma_{\rm
  v}=417$~km~s$^{-1}$ derived by MZ. Despite this relatively large
velocity dispersion, the smoothed image in Fig.~\ref{fig,smoothed}
shows no evidence for a concentration of emission around the
luminosity-weighted optical group centre at ($12^{\rm h} 35^{\rm m}
01\fs 20$, $-03\degr 37\arcmin 09\farcs 3$).  The surface brightness
profile in Fig.~\ref{fig,surfbright}, centred on this position due to
the lack of any obvious X-ray peak, does not exhibit any systematic
radial trend either, nor does it suggest the presence of significant
excess emission at any radius. The X-ray maps confirm the absence of
emission above the background within both 200~kpc and $0.5r_{\rm vir}$
from the optical group centre, and we conclude that no emission is
detected at 90~per~cent confidence outside individual galaxies.
 
The $\sigma_{\rm v}$--$T$ relation of \citet{osmo04} would suggest
$T=1.1^{+0.2}_{-0.1}$~keV
for the IGM in this group.  Our failure to detect any IGM emission
then implies a $3\sigma$ upper limit to the diffuse 0.3--2~keV
luminosity of $L_{\rm X} < 3.3 \times 10^{41}$~erg~s$^{-1}$ inside
$0.5r_{\rm vir}$ for any subsolar metallicity.  Again, this can be
compared to the expectation $L_X\sim 4\times 10^{42}$~erg~s$^{-1}$
from the $L_{\rm X}$--$\sigma_{\rm v}$ relation of \citet{osmo04}.
Like MZ~4577, this group is thus considerably X-ray fainter than
expected from the velocity dispersion of its galaxies. For the assumed
limits on $T$ and $Z$, the luminosity limit implies $\langle n_e
\rangle < 3.2\times 10^{-5}$~cm$^{-3}$, $M_{\rm gas} < 2.2\times
10^{12}$~M$_\odot$, and $\langle t_{\rm cool} \rangle >2.9\times
10^{11}$~yr inside $0.5r_{\rm vir}$.
 
32 point sources are detected in the field, down to $\sim 3 \times
10^{-15}$~erg~cm$^{-2}$~s$^{-1}$. Emission (inside 0.5~arcmin) at a
level of $\sim 200$ net counts is seen around the first and
second-ranked galaxies (both ellipticals).  Their spectra suggest
thermal plasmas with $T = 0.7\pm 0.1$ and $0.6\pm 0.1$~keV, and
unabsorbed 0.3--2~keV luminosities of $L_{\rm X} = 2.8\pm 0.7\times
10^{41}$ and $2.3\pm 0.3\times 10^{41}$~erg~s$^{-1}$, respectively.
Useful constraints on their gas metallicities could not be obtained,
so $Z=0.3$~Z$_\odot$ has been assumed here. Unless counteracted by
feedback, this hot gas should be able to cool efficiently within the
haloes of the galaxies, having mean cooling times of only
$3.2^{+0.5}_{-0.3}$ and $2.5^{+0.2}_{-0.1}$~Gyr.

\subsection{MZ~9014}\label{sec,mz9014}

After flare cleaning, this is the deepest X-ray observation within the 
present sample. The updated velocity dispersion of the 22 
spectroscopically confirmed member galaxies is consistent with the 
earlier value of $\sigma_{\rm v} = 240$~km~s$^{-1}$ from the 
MZ~catalogue. 
 
As can be discerned from Fig.~\ref{fig,smoothed}, faint, irregular IGM
emission is detected both within 200~kpc and $0.5r_{\rm vir}$ at more
than $3\sigma$~significance.  This emission is not peaking at the
optical group centre which is in fact slightly X-ray fainter than its
immediate surroundings (cf.\ Fig.~\ref{fig,smoothed}).  Exhibiting
only $\sim 170$ net counts, the emission is concentrated along a broad
ridge, roughly coinciding with the region occupied by 17 of the 22
member galaxies. Given the absence of a clear X-ray peak and the fact
that the brightest group galaxy is located just on the eastern edge of
the detected diffuse emission (Fig.~\ref{fig,X-opt}), the surface
brightness profile shown in Fig.~\ref{fig,surfbright} has been centred
on the diffuse emission centroid at ($00^{\rm h} 37^{\rm m} 40\fs 70$,
$-27\degr 30\arcmin 31\farcs 1$). This was evaluated from an
unsmoothed version of Fig.~\ref{fig,smoothed} inside a circle
enclosing the outermost contour of the 'ridge' seen in
Fig.~\ref{fig,X-opt}. A spectral analysis was also attempted, but a
thermal plasma model fit leaves both temperature and abundance
unconstrained at 90~per~cent confidence.  However, within the
$1\sigma$ errors, the nominal best-fitting value of $T = 0.6\pm
0.3$~keV is independent of the choice of any subsolar value of $Z$,
and is furthermore in good agreement with the value of $T = 0.6\pm
0.2$~keV
suggested by the $\sigma_{\rm v}$--$T$ relation of \citet{osmo04}.
 
These results imply a very low IGM luminosity of $L_{\rm X}=6\pm 2
\times 10^{40}$~erg~s$^{-1}$ inside $0.5r_{\rm vir}$ for any subsolar
metallicity, an order of magnitude below the expectation $L_{\rm X}
\approx 1\times 10^{42}$~erg~s$^{-1}$ from both the $L_{\rm
  X}$--$\sigma_{\rm v}$ and the $L_{\rm X}$--$T$ relation.  The
derived $L_{\rm X}$ is lower than that of any system with detectable
IGM emission in the {\em ROSAT}--based sample of \citet{osmo04},
demonstrating the superior ability of {\em XMM} to detect low surface
brightness emission.  For the hot gas in the group, we find
corresponding ranges of $\langle n_e \rangle = 1.0-3.7\times
10^{-5}$~cm$^{-3}$, $M_{\rm gas} = 2.8-10.4\times 10^{11}$~M$_\odot$,
and $\langle t_{\rm cool} \rangle$ in the range 1.1--$6.8 \times
10^{10}$~yr, for any $Z\leq $~Z$_\odot$.
 
Despite detecting 74 point sources in the field down to $\sim 2\times
10^{-15}$~erg~cm$^{-2}$~s$^{-1}$, only the optically second-brightest
group galaxy, an elliptical, is picked up in X-rays, at a level of
$\sim 70$ net counts.  Binning its spectrum into 5 counts per channel,
and fitting a $Z=0.3$~Z$_\odot$ {\sc mekal} model using Cash
statistics, yields $T=0.4\pm 0.1$~keV and $L_{\rm X} =
4.0^{+1.5}_{-0.8}\times 10^{40}$~erg~s$^{-1}$, with $\langle t_{\rm
  cool}\rangle = 3.8\pm 0.5$~Gyr.  The brightest group galaxy, a
spiral, is not detected, implying $L_{\rm X}<1.7\times
10^{40}$~erg~s$^{-1}$.

\subsection{MZ~9307}

This is another high-$\sigma_{\rm v}$ group, along with MZ~5383 the
only system in our sample to have $\sigma_{\rm v} > 400$~km~s$^{-1}$
in the original MZ~catalogue. The {\em XMM} observation was performed
using the medium optical blocking filter due to a bright star in the
field.  The presence of significant background flares in the X-ray
data did not preclude valuable constraints on the diffuse emission
level to be be obtained.
 
No diffuse emission is seen in the X-ray maps, and the surface
brightness profile, centred on the luminosity-weighted group centre at
($00^{\rm h} 40^{\rm m} 47\fs 11$, $-27\degr 28\arcmin 18\farcs 8$),
confirms the lack of any enhancement of emission both inside 200~kpc
and $0.5r_{\rm vir}$.  The dip in the profile seen around $r\approx
40$~arcsec is present also in the raw data and is thus not an artefact
of the background subtraction or exposure correction. The velocity
dispersion would suggest $T=1.0\pm 0.3$~keV for any hot group gas, and
the lack of detectable IGM emission then translates into a $3\sigma$
upper limit of $L_{\rm X} < 5.1\times 10^{41}$~erg~s$^{-1}$ for $Z\leq
$~Z$_\odot$, with $\langle n_e \rangle < 8.0\times 10^{-5}$~cm$^{-3}$,
$M_{\rm gas} < 1.5\times 10^{12}$~M$_\odot$, and $\langle t_{\rm cool}
\rangle >1.0\times 10^{11}$~yr inside $0.5r_{\rm vir}$.
 
We detect 23 point sources in the field, down to a limiting flux of 
$\sim 2\times 10^{-15}$~erg~cm$^{-2}$~s$^{-1}$.  One of these is an 
X-ray bright background quasar at $z = 0.170$, just visible to the 
upper right in Fig.~\ref{fig,smoothed}.  While none of the seven 2dF 
galaxies in the group is picked up as an X-ray source, our forthcoming 
IMACS spectroscopy will address whether any optically fainter group 
galaxies have X-ray counterparts.

\begin{table*} 
 \centering 
 \begin{minipage}{103mm} 
  \caption{X-ray properties of the IGM and the optically brightest 
    group galaxies, down to the brightest X-ray undetected galaxy. Column~3 
    specifies whether the galaxy is of early (E) or late (S) type.} 
  \label{tab,results} 
  \begin{tabular}{@{}lccccrr@{}} \hline \multicolumn{1}{l}{Group} &
    \multicolumn{1}{c}{Rank} & \multicolumn{1}{c}{E/S?} &
    \multicolumn{1}{c}{$M_B$} & \multicolumn{1}{c}{$T_{\rm X}$} &
    \multicolumn{1}{c}{$L_{\rm X}$} &
    \multicolumn{1}{c}{$\langle t_{\rm cool}\rangle$}\\
    & & & & (keV) & ($10^{40}$~erg~s$^{-1}$) & (Gyr)$\ $ \\ \hline
    MZ~4577 & IGM & \ldots & \ldots & $0.5^{+0.4}_{-0.1}$\footnote{Value
      derived assuming $T_{\rm X}$ from the $\sigma_{\rm V}$--$T_{\rm
        X}$ relation of \citet{osmo04}, with $\sigma_{\rm V}$ from
      Table~\ref{tab,opt} (see Section~\ref{sec,obs} for details).} & $21\pm 10^a$ & 24--97$^a$ \\
    \ldots  & 1   & S      & $-20.45$  & \ldots & $<2.7$ & \ldots \\
    MZ~5383 & IGM & \ldots & \ldots & $1.1^{+0.2}_{-0.1}$ $^a$ & $<33^a$  & $>290^a$ \\
    \ldots  & 1   & E & $-21.13$   & $0.7\pm 0.1$ & $28\pm 7$  & $3.2^{+0.5}_{-0.3}$ \\
    \ldots  & 2   & E & $-20.87$   & $0.6\pm 0.1$ & $23\pm 3$  & $2.5^{+0.2}_{-0.1}$ \\
    \ldots  & 3   & S & $-20.38$   & \ldots & $<3.3$ & \ldots \\
    MZ~9014 & IGM & \ldots & \ldots & $0.6\pm 0.3$ & $6\pm 2$ & 11--68 \\
    \ldots  & 1 & S & $-20.90$ & \ldots & $<1.7$ & \ldots  \\
    \ldots  & 2 & E & $-20.86$ & $0.4\pm 0.1$ & $4.0^{+1.5}_{-0.8}$ & $3.8\pm 0.5$ \\
    MZ~9307 & IGM & \ldots & \ldots & $1.0\pm 0.3^a$ & $<51^a$ & $>100^a$ \\
    &  1  &  E     & $-18.87 $  & \ldots & $<2.1$ & \ldots \\
    \hline
\end{tabular} 
\end{minipage} 
\end{table*}

\section{Discussion}\label{sec,discuss} 
 
\subsection{Comparison to X-ray selected groups}\label{sec,comparison} 
 
Summarizing the X-ray properties of the hot IGM in the four groups, we
find that two groups, MZ~5383 and MZ~9307, show no detectable diffuse
emission, whereas MZ~4577 shows some evidence of IGM emission though
this is not significant at the $3\sigma$ level. Only in one system,
MZ~9014, do we detect some irregular emission at
$>3\sigma$~significance. The X-ray luminosity of this system is among
the lowest found for any X-ray detected group, and since its hot gas
is clearly not in hydrostatic equilibrium, no X-ray mass analysis was
attempted.
 
On the basis of their velocity dispersions, all four groups are
remarkably X-ray underluminous with respect to typical X-ray selected
groups, as illustrated in Fig.~\ref{fig,lxsig}. This plot shows the
corresponding results from the GEMS sample of \citet{osmo04}, along
with several observationally derived $L_{\rm X}$--$\sigma_{\rm V}$
relations spanning a large range of derived slopes, from $L_{\rm X}
\propto \sigma_{\rm V}^{1.56}$ \citep{mahd97} to $L_{\rm X} \propto
\sigma_{\rm V}^{4.5}$ \citep{hels00} (all normalized to our adopted
value of $H_0$). Even when comparing to the combined results from
these highly disparate data sets, the {\em XI} groups emerge as
significantly X-ray underluminous for their velocity dispersions. This
is particularly true in light of the fact that, apart from the
\citet{osmo04} relation, the X-ray luminosities underlying the $L_{\rm
  X}$--$\sigma_{\rm V}$ relations shown in Fig.~\ref{fig,lxsig} have
generally not been extrapolated to $r_{500}$. Doing so would raise the
normalization of the relations,
thus aggravating the discrepancy with respect to the {\em XI} groups.
It is perhaps also surprising that it is the two {\em XI} groups with
the largest values of $\sigma_{\rm v}$ that remain undetected in the
X-ray.  These results indicate already that the full {\em XI} sample
will have properties quite different from the heterogeneous group
samples studied in the past, and that these earlier samples provided a
biased view of the group population.
 
\begin{figure} 
 \mbox{\hspace{-6mm} 
 \includegraphics[width=94mm]{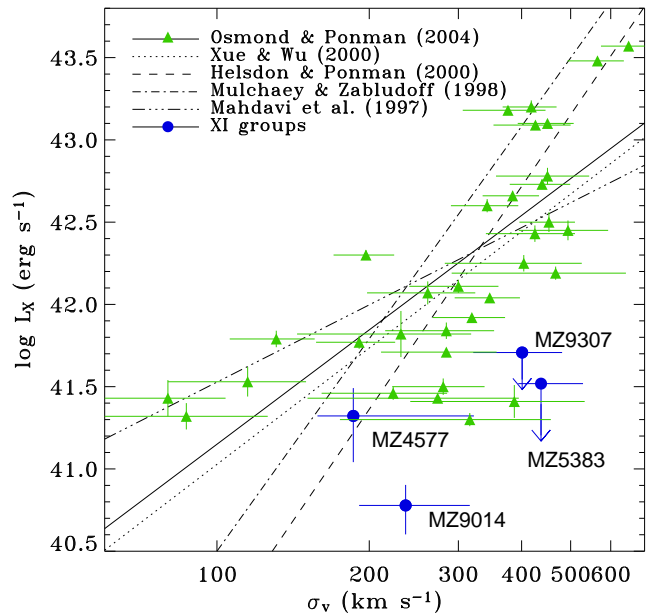}} 
\caption{$L_{\rm X}$--$\sigma_{\rm v}$ relations for X-ray bright 
  groups. Errors on $L_{\rm X}$ for the {\em XI} groups are 
  90~per~cent confidence, with downward arrows representing $3\sigma$ 
  upper limits.  Overplotted for comparison are the data points of 
  \citet{osmo04}.} 
\label{fig,lxsig} 
\end{figure}

Given the importance of this conclusion, it seems worth addressing the 
robustness of the result shown in Fig.~\ref{fig,lxsig}, before 
proceeding to investigate its source of origin.  One potential concern 
is that $L_{\rm X}$ of our groups has been evaluated inside a fiducial 
radius of $0.5r_{\rm vir}$ (assumed roughly equal to $r_{500}$), for 
physical consistency and straightforward comparison to the results of 
\citet{osmo04}. The reliability of the virial radii adopted from MZ 
for this purpose could be questioned, as these are in some cases based 
on only five galaxies. This issue will be addressed in more detail 
when improved membership statistics become available for a larger 
sample of {\em XI} groups, but we note for now that the main 
conclusion -- that our four groups are X-ray underluminous for their 
velocity dispersions -- is clearly robust to reasonable changes in 
$r_{\rm vir}$. 
 
Another issue is whether the velocity dispersions could be
overestimated in our analysis. We tested this for the three groups
with available IMACS velocities, by estimating errors on galaxy
velocities both from the template fitting code and from the variation
in derived redshifts resulting from manually fitting multiple line
features. Reassuringly, the resulting values of $\sigma_{\rm v}$ were
entirely consistent.  The argument that $\sigma_{\rm v}$ could be
artificially boosted by contamination from interlopers could possibly
be justified for a single group, but hardly for all four, and
certainly not by a factor of 2--3 required to bring MZ~5383, MZ~9014,
and MZ~9307 into formal agreement with the $L_{\rm X}$--$\sigma_{\rm
  v}$ relation of \citet{osmo04}.  For MZ~9307, the velocity
dispersion adopted from MZ could potentially be unreliable due to the
small number of galaxies involved.  However, MZ derive $\sigma_{\rm
  v}$ for groups with $N_{\rm gal}\leq 15$ using the gapper estimator
(see e.g.\ \citealt{beer90}), which should improve robustness in
$\sigma_{\rm v}$--estimates for poorly sampled systems. The
consistency of the IMACS results for $\sigma_{\rm v}$ with the
corresponding MZ values for the three other groups also suggests that
the adopted velocity dispersion of MZ~9307 cannot be seriously biased.

\subsection{Physical state of the IGM} 
 
With the robustness of the result shown in Fig.~\ref{fig,lxsig} 
reasonably well established, the question remains how to interpret the 
lack of significant X-ray emission in our groups, and whether this 
lack of emission is coupled to their dynamical state.  It is most 
unlikely that the two undetected groups are not gravitationally bound, 
since (as mentioned in Section~\ref{sec,sample}) only groups with 
number density contrasts $\delta \rho/ \langle \rho \rangle \geq 80$ 
have been included in the MZ catalogue. This leaves at least three 
possible explanations for the lack of significant IGM emission in our 
sample. 
 
(i) The groups could be in the process of collapsing for the first
time, in which case the present velocity dispersion could be a poor
proxy of the depth of the gravitational potential of the group or the
temperature of its X-ray gas. This scenario allows for significant
amounts of intragroup gas, which, however, would be more uniformly
distributed than in collapsed groups and may not yet have been
shock-heated to the virial temperature of the final system.  In this
case one might also expect to see evidence of dynamical substructure
in the galaxy distribution.  To investigate this, we plot velocity
histograms of the groups in Fig.~\ref{fig,velhist}. There is generally
some indication, particularly for MZ~5383, of a central gap in the
histogram. The result for MZ~4577 also suggests a bimodal velocity
distribution, with a subgroup of four galaxies separated from the
remaining nine galaxies by $\sim 300$~km~s$^{-1}$. These features
could be associated with the merging of two subgroups, but there is no
convincing evidence that any apparent velocity substructures are
localized on the sky in any of the groups. However, given the fairly
modest number of galaxies found in each group, it is likely that
general conclusions about velocity substructure in these four groups
cannot be reached on the basis of individual velocity histograms.  To
verify this, we confirmed, using the data from \citet{zabl98}, that
apparently virialised systems with a comparable number of confirmed
members, such as HCG~42 and NGC~4325, can show velocity histograms
with features similar to those seen in Fig.~\ref{fig,velhist}.

\begin{figure} 
 \mbox{\hspace{0mm} 
 \includegraphics[width=86mm]{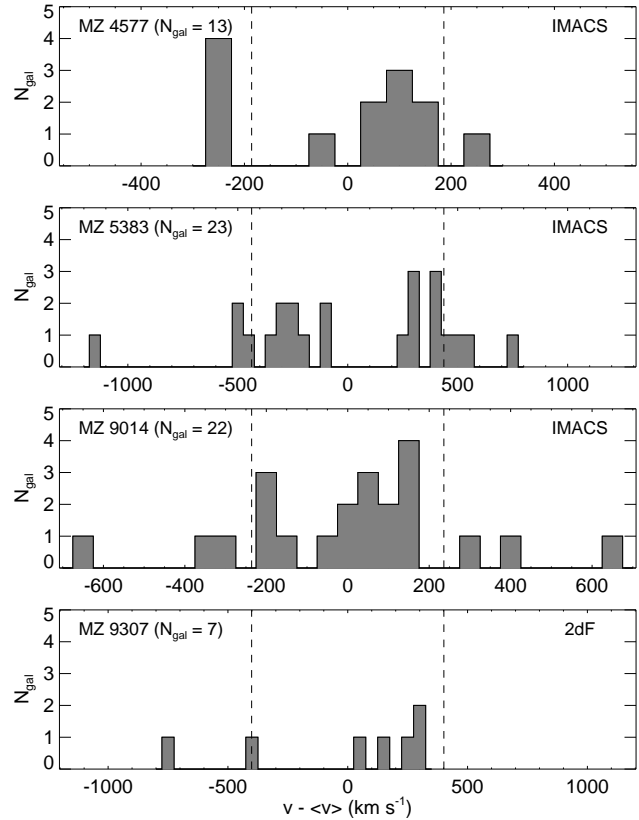}} 
\caption{Histogram of galaxy velocities relative to the mean velocity 
  for each group in 50~km~s$^{-1}$~bins across the velocity range $\pm 
  3\sigma_{\rm v}$. Dashed lines mark $\sigma_{\rm v}$ in each case.} 
\label{fig,velhist} 
\end{figure} 

Potentially, some insight into the dynamics of the groups can still be
gained by combining all the velocity measurements. To this end, we
constructed a 'pseudo-group' containing all confirmed member galaxies,
by normalizing the galaxy velocities shown in Fig.~\ref{fig,velhist}
to the velocity dispersion of the relevant group.
Fig.~\ref{fig,velhist_norm} shows the resulting stacked histogram of
the 65 normalized galaxy velocities. It is clear that the central
deficit of galaxy velocities persists in this representation.  Using a
K--S test, we find that the probability that the normalized velocities
have been drawn from a Gaussian with unit variance $\sigma_{\ast}^2 =
[(v-\langle v\rangle)/\sigma_{\rm V}]^2$, is $P=0.093$ for all 65
galaxies, decreasing to $P=0.068$ if including only the 58 galaxies
with IMACS velocity measurements. Although still dealing with a
relatively small number of galaxies, this result may suggest that the
groups, as a class, are still in the process of dynamical relaxation.

\begin{figure} 
 \mbox{\hspace{-2mm} 
 \includegraphics[width=86mm]{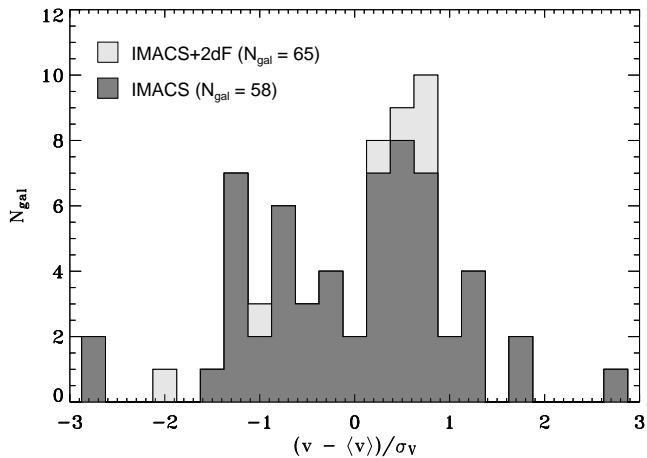}} 
\caption{Stacked histogram of all galaxy velocities, measured relative to
  the mean velocity for each group and normalized to the velocity
  dispersion of that group.  Dark shaded area outlines the histogram
  for the three groups with IMACS velocities, while the lighter shaded
  area also includes MZ~9307.}
\label{fig,velhist_norm} 
\end{figure}

A more detailed dynamical analysis of the {\em XI} sample will be
undertaken when results for the full sample become available. Within
the framework of the present analysis, we note that a picture in which
the groups are still collapsing also has support from studies of other
group samples. These include both purely optical analyses (e.g.\
\citealt{tull87}; \citealt{gira00} and references therein) and
X-ray/optical comparisons (e.g.\ \citealt{pope05}). In particular,
\citet{gira00} infer a limiting number density contrast of $\sim 70$
for their 'P' group subsample, comparable to that of the MZ catalogue,
and estimate that most of these groups are in the phase of collapse.
\citet{pope05} investigated RASS data for a sample of
spectroscopically confirmed Abell clusters and found that 50 out of
138 clusters were X-ray underluminous for their optically derived
virial mass. Although detecting no obvious optical substructure in
their individual X-ray faint clusters, their superior statistics
enabled these authors to conclude, on the basis of the position and
velocity distributions of the cluster members, that these systems as a
class are most likely still collapsing. Further support for ongoing
collapse of a considerable fraction of the group population comes from
cosmological simulations, which suggest that 30--50 per cent of all
groups selected via standard FOF-algorithms could be collapsing for
the first time at the present epoch (J.~Sommer-Larsen, priv.\ comm.).

(ii) A second explanation for the general lack of significant X-ray
emission in the groups is that the gravitational potentials of these
groups are too shallow to heat the intragroup gas to X-ray
temperatures.  As in scenario (i) above, there could be plenty of
intragroup gas, but in this case its temperature would largely remain
too low ($\la 10^6$~K) to render it detectable in these observations.
However, given the high values measured for the velocity dispersions
of our groups, this explanation seems unattractive, and, as has
already been discussed, the velocity dispersions of our groups are
unlikely to be substantially overestimated. The fact that we {\em do}
detect hot gas in two of the groups also suggests that this scenario
cannot offer an exhaustive explanation, even if assuming that the
measured velocity dispersions are poor proxies of the total group
mass.  The long mean cooling times derived for the X-ray detected gas
indicate that the gas is emitting inefficiently in the X-ray band due
to its density, rather than temperature, being surprisingly low. The
fact that MZ~9014 is underluminous relative to the expectation from
the $L_{\rm X}$--$T$ relation but is consistent with the $\sigma_{\rm
  V}$--$T$ relation (cf.\ Section~\ref{sec,mz9014}), seems to support
this interpretation.

(iii) Alternatively, the groups could be X-ray faint because many
collapsed groups simply contain very little intragroup gas. Our
unbiased selection could thus be picking up such systems because they
are more numerous than X-ray bright groups. One mechanism which could
give rise to this situation is strong galactic feedback ejecting a
significant fraction of the original IGM from the group potential.
However, that would leave the challenge of explaining why feedback
would be so much stronger in some groups as to reduce the X-ray
detectable hot gas mass by 1--2 orders of magnitude relative to other
systems with potential wells of similar depth.  Moreover, since the
formation of elliptical galaxies is expected to generate more feedback
than that of spirals \citep{arna92}, strong group-wide feedback would
probably require a galaxy population with a low spiral fraction
$f_{\rm sp} = N_{\rm sp}/N_{\rm gal}$. If tentatively identifying
those 2dF group galaxies with values of the spectral type parameter
$\eta > -1.4$ as spirals (see \citealt{madg02}), then the mean spiral
fraction of our four groups is $\sim 65$~per~cent. Such a large value
of $f_{\rm sp}$ may not be easily reconciled with the notion that
galactic feedback has ejected much of the IGM from the groups.
 
We note that \citet{mahd00} detected X-ray emission in RASS data from
only 42 out of a statistically complete sample of 260 groups,
suggesting that $\sim 75$~per~cent of their groups do not contain a
hot IGM.  However, due to the very shallow RASS exposures, these
authors could not study X-ray emission from groups with $L_{\rm X}\la
10^{42}$~erg~s$^{-1}$, which would be expected to be more common. In
fact, based on the expectation from X-ray bright groups, more than
half of the 25 {\em XI} groups are expected to show X-ray emission
around or below this limit (cf.\ Figs.~\ref{fig,sigmarv} and
\ref{fig,lxsig}).  Until a larger sample of such groups has been
studied using more sensitive X-ray data, it seems premature to accept
scenario (iii) without considering viable alternatives.

In summary, we cannot at this stage exclude the possibility that the
surprisingly low diffuse X-ray luminosity of our groups is due to any
IGM being either largely absent or too cold to produce copious X-ray
emission.  In order to help constrain the {\em total} gas content of
the {\em XI} groups, we have commenced a programme of H{\sc i} imaging
of the groups using the Giant Metrewave Radio Telescope.  When H{\sc
  i} results for a significant number of groups become available, we
will be in a better position to assess the validity of scenarios ii)
and iii) above.  Until then, we find the idea that the groups could
still be in the process of virialisation more attractive.  In contrast
to the other scenarios discussed, this explanation draws support from
both cosmological simulations and other observational group studies,
without facing any immediate challenges.

We note that the morphological composition and the likely dynamical
status of these groups suggest an analogy to the Local Group (LG) of
galaxies, which consists of three spirals brighter than
$M_V\!\sim\!-19$, with a varied assortment of early and late-type
dwarfs, and is collapsing for the first time.  Although having a low
velocity dispersion of $\sigma_{\rm v}\sim 100$~km~s$^{-1}$,
comparable to those of the lowest-$\sigma_{\rm v}$ groups in the {\em
  XI} sample (Fig.~\ref{fig,sigmarv}), the total LG mass of $\sim
2\times 10^{12}$~M$_\odot$ (e.g.\ \citealt{vand99}) suggests a mass
overdensity with respect to the critical density of $\sim 45$, within
the region occupied by the Milky Way and M31 subgroups.  The
equivalent galaxy number density contrast in the adopted cosmology of
$\delta \rho/\langle \rho \rangle \sim 150$ suggests that the LG would
in fact meet the overdensity criterion for inclusion in the MZ
catalogue.  Though no X-ray emitting hot IGM has yet been detected in
the LG, which is unlikely to have a diffuse X-ray luminosity of more
than a few times $10^{40}$~erg~s$^{-1}$ \citep{rasm01}, recent
detections of zero-redshift O{\sc vi} and O{\sc vii} absorption lines
in {\em FUSE} and {\em Chandra} observations of quasars suggest the
presence of an LG intergalactic medium of electron density $\sim
5\times 10^{-6}$~cm$^{-3}$ and temperature $T\la 10^6$~K (e.g.\
\citealt{nica03}). This is consistent with the low fluxes we observe
for our groups. Our H{\sc i} studies of the {\em XI} groups will help
us compare their cold gas content with that of the LG.

\subsection{Properties of group galaxies} 
 
As will be described in more detail in a subsequent paper, a
substantial fraction of the {\em XI} group members are found from our
IMACS spectroscopy to be emission-line galaxies. In particular, strong
optical emission lines are found in roughly half of the confirmed
member galaxies in the three groups with IMACS data discussed here.
Hence, star-forming galaxies and active galactic nuclei are seen in
all of these groups.
Given this, it is perhaps surprising that out of a total of 58
confirmed group galaxies, we detect X-ray emission above a few times
$10^{40}$~erg~s$^{-1}$ in only three of them (cf.\
Tables~\ref{tab,opt} and \ref{tab,results}). These three galaxies are
all ellipticals, displaying no evidence for significant optical
emission lines. Their X-ray luminosities are fairly typical of
moderately X-ray bright ellipticals, and their measured luminosities
and temperatures compare well to the expectations from the $L_{\rm
  X}$--$T$ relation derived for ellipticals by \citet{osul03}.  The
galaxies that show AGN-like optical emission in our groups must have
X-ray emission below our detection limit of $\sim 2-3 \times
10^{40}$~erg~s$^{-1}$, and their nuclei could be related to the class
of low-luminosity AGN seen in some nearby galaxies (see, e.g.,
\citealt{tera02}).

From an optically selected sample of groups having pointed {\em ROSAT}
PSPC coverage, \citet{zabl98} found that groups with detectable X-ray
emission typically have a bright ($M_B\la M_B^{\ast}-1\approx -21.4$)
elliptical at the centre of the X-ray emission.  In addition to the
unexpectedly faint IGM emission of the four {\em XI} groups studied
here, another potential difference between these and more X-ray
luminous systems is therefore that none of our groups hosts a dominant
central elliptical galaxy.  In two of the groups, including the X-ray
detected MZ~9014, the optically most luminous galaxy is in fact a
spiral. None of the ellipticals is 1~mag brighter than $M_B^{\ast}$,
and we note in particular that the brightest 2dF galaxy in MZ~9307 is
$\sim 1.5$~mag {\em fainter} than $M_B^{\ast}$.  The absence of a
dominant elliptical in the groups also seems to support the hypothesis
that the groups are still collapsing, since such galaxies are probably
formed via mergers in dense environments.
 
We should also note, though, that the two X-ray detected groups,
MZ~4577 and MZ~9014, are not the only systems known to show diffuse
intragroup X-ray emission which is not peaking on a central, bright
early-type galaxy.  Other examples include the spiral-dominated Hickson
compact groups HCG~16 \citep{doss99,bels03}, HCG~57 \citep{fuka02},
and the well-studied HCG~92 (Stephan's Quintet;
\citealt*{sule95,awak97,piet97,trin03,trin05}), as well as the
early-type dominated HCG~37 \citep{mulc03}. These groups all exhibit
low diffuse X-ray luminosities in the range $\approx$~1--5$\times
10^{41}$~erg~s$^{-1}$, comparable to $L_{\rm X}$ of the X-ray detected
{\em XI} groups. Systems like the Hickson groups, likely to represent
groups close to maximum collapse, are not very common in the nearby
Universe, however. For example, we find that only about $\sim
1$~per~cent of the groups in the MZ catalogue (and, as mentioned in
Section~\ref{sec,sample}, none of the {\em XI} groups) satisfy the
\citet{hick82} criteria.  Outside Hickson's catalogue, other examples
are NGC~7777 and SHK~202 \citep{mulc03}.  However, the number of such
systems in the literature is still very low, placing MZ~4577 and
MZ~9014 in an exclusive club.  Coupled with the results described in
Section~\ref{sec,comparison}, this is a further indication that the
{\em XI} project is targeting a class of groups not previously studied
in much detail in X-rays.

\section{Conclusions} 
 
As part of an ongoing effort to investigate the X-ray and optical
properties of a substantial, statistically unbiased, and kinematically
selected sample of galaxy groups, for the first time using deep X-ray
data, we have performed {\em XMM-Newton} observations of the first
four groups in this sample. In two of the groups, we detect an X-ray
emitting intragroup medium, with luminosities among the lowest found
for X-ray detected groups so far. The two other groups observed here
remain undetected in the X-ray, and all four groups are found to be
X-ray underluminous for their velocity dispersions when compared to
expectations from X-ray bright groups. Furthermore, none of the groups
hosts a dominant elliptical galaxy at the centre of the X-ray
emission.  Our results therefore suggest that the nature of the IGM in
these optically selected groups may be very different from that seen
in standard X-ray selected group samples. The fact that we are finding
some of the faintest IGM yet seen in groups in our first four
observations clearly demonstrates that the {\em XI} Project is
exploring new parameter space. It also establishes that the
predominantly X-ray selected group samples studied in the past with
e.g.\ {\em ROSAT} were not representative of the overall group
population.  In contrast, the unbiased redshift selection employed in
the {\em XI} Project appears to be targeting a different class of
groups not previously studied in much detail.
 
The low levels of IGM emission in the groups could be an indication
that i) the groups are in the process of collapsing for the first
time, ii) the gravitational potentials are too shallow for the gas to
emit much X-ray emission (i.e.\ the gas is essentially too cool to
produce X-rays), or iii) there is simply little or no intragroup gas.
We find the first explanation to be the more attractive, at least
until the validity of the other scenarios can be more firmly addressed
by subjecting a larger sample of optically selected groups to deep
X-ray observations.  The idea that many groups are collapsing for the
first time at the present epoch is also consistent with expectations
from other optical group studies and with results of cosmological
simulations.
 
In contrast to the more well-studied class of X-ray bright groups, the
type of groups studied in this paper truly represents the most common
galaxy environments in the Universe. Investigating the detailed
properties of a larger sample of such groups is therefore crucial if
one is to obtain an unbiased understanding of the nature of the group
population and a census of the distribution of baryons in the
Universe.

\section*{Acknowledgments} 
We thank Gus Oemler, Dan Kelson, and Greg Walth for their role in
developing the IMACS reduction pipeline, Gary Mamon, Andrea Biviano,
and Jesper Sommer-Larsen for useful discussions, and the referee for
comments which helped improve this paper. JR acknowledges the support
of the European Community through a Marie Curie Intra-European
Fellowship under contract no.\ MEIF-CT-2005-011171.  JSM acknowledges
partial support from NASA grants NNG04GF78G and NNG04GC846.

\bsp 
 
\label{lastpage} 
 

\begin{thebibliography}{} 
\bibitem[\protect\citeauthoryear{Arnaud et al.}{1992}]{arna92}  
  Arnaud M., Rothenflug R., Boulade O., Vigroux L., Vangioni-Flam E., 1992,  
  A\&A, 254, 49  
\bibitem[\protect\citeauthoryear{Arnaud et al.}{2002}]{arna02}  
  Arnaud M.\ et al., 2002, A\&A, 390, 27  
\bibitem[\protect\citeauthoryear{Awaki et al.}{1997}]{awak97}  
  Awaki H., Koyama K., Matsumoto H., Tomida H., Tsuru T., Ueno S., 1997, PASJ,
  49, 445  
\bibitem[\protect\citeauthoryear{Barkhouse et al.}{2006}]{bark06} 
  Barkhouse W.A., et al., 2006, ApJ, 645, 955
\bibitem[\protect\citeauthoryear{Beers, Flynn \& Gebhardt} 
  {Beers et al.}{1990}]{beer90}  
  Beers T.C., Flynn K., Gebhardt K., 1990, AJ, 100, 32  
\bibitem[\protect\citeauthoryear{Belsole et al.}{2003}]{bels03} 
  Belsole E., Sauvageot J.-L., Ponman T.J., Bourdin H., 2003, A\&A, 398, 1 
\bibitem[\protect\citeauthoryear{Bertin \& Arnouts}{1996}]{bert96} 
  Bertin E., Arnouts S., 1996, A\&AS, 117, 393 
\bibitem[\protect\citeauthoryear{Bigelow \& Dressler}{2003}]{bige03} 
  Bigelow B.C., Dressler A.M., 2003, Proc.\ SPIE, 4841, 1727  
\bibitem[\protect\citeauthoryear{Bullock et al.}{2001}]{bull01}  
  Bullock J.S., Kolatt T.S., Sigad Y., Somerville R.S., Kravtsov A.V., Klypin  
  A.A., Primack J.R., Dekel A., 2001, MNRAS, 321, 559  
\bibitem[\protect\citeauthoryear{Burns et al.}{1996}]{burn96}  
  Burns J.O., Ledlow M.J., Loken C., Klypin A., Voges W., Bryan G.L., Norman  
  M.L., White R.A., 1996, ApJL, 467, L49  
\bibitem[\protect\citeauthoryear{Colless et al.}{2001}]{coll01}  
  Colless M.\ et al., 2001, MNRAS, 328, 1039  
\bibitem[\protect\citeauthoryear{De Propris et al.}{2002}]{depr02}  
  De Propris R.\ et al., 2002, MNRAS, 329, 87  
\bibitem[\protect\citeauthoryear{Donahue et al.}{2002}]{dona02}  
  Donahue M.\ et al., 2002, ApJ, 569, 689  
\bibitem[\protect\citeauthoryear{Dos Santos \& Mamon}{1999}]{doss99} 
  Dos Santos S., Mamon G.A., 1999, A\&A, 352, 1  
\bibitem[\protect\citeauthoryear{Ebeling, Voges \& B\"{o}hringer} 
  {Ebeling et al.}{1994}]{ebel94}  
  Ebeling H., Voges W., B\"{o}hringer H., 1994, ApJ, 436, 44  
\bibitem[\protect\citeauthoryear{Eke, Cole \& Frenk} 
  {Eke et al.}{1996}]{eke96}  
  Eke V.R., Cole S., Frenk, C.S., 1996, MNRAS, 282, 263  
\bibitem[\protect\citeauthoryear{Eke et al.}{2004}]{eke04}  
  Eke V.R.\ et al., 2004, MNRAS, 348, 866  
\bibitem[\protect\citeauthoryear{Fukazawa et al.}{2002}]{fuka02}  
  Fukazawa Y., Kawano N., Ohto A., Mizusawa H., 2002, PASJ, 54, 527  
\bibitem[\protect\citeauthoryear{Fukugita, Hogan \& Peebles} 
  {Fukugita et al.}{1998}]{fuku98}  
  Fukugita M., Hogan C.J., Peebles P.J.E., 1998, ApJ, 503, 518  
\bibitem[\protect\citeauthoryear{Gilbank et al.}{2004}]{gilb04}  
  Gilbank, D.G., Bower R.G., Castander F.J., Ziegler B.L., 2004, MNRAS, 348,  
  551  
\bibitem[\protect\citeauthoryear{Girardi \& Giuricin}{2000}]{gira00} 
  Girardi M., Giuricin G., 2000, ApJ, 540, 45  
\bibitem[\protect\citeauthoryear{Helsdon \& Ponman}{2000}]{hels00} 
  Helsdon S.F., Ponman, T.J., 2000, MNRAS, 315, 356 
\bibitem[\protect\citeauthoryear{Helsdon \& Ponman}{2003}]{hels03} 
  Helsdon S.F., Ponman, T.J., 2003, MNRAS, 340, 485 
\bibitem[\protect\citeauthoryear{Hicks et al.}{2004}]{hick04}  
  Hicks A., Ellingson E., Bautz M., Yee H., Gladders M., Garmire G., 2004,  
  35th COSPAR Scientific Assembly, 1624  
\bibitem[\protect\citeauthoryear{Hickson}{1982}]{hick82} 
  Hickson P., 1982, ApJ, 255, 382  
\bibitem[\protect\citeauthoryear{Kelson}{2003}]{kels03} 
  Kelson D.D., 2003, PASP, 115, 688 
\bibitem[\protect\citeauthoryear{Ledlow et al.}{1996}]{ledl96}  
  Ledlow M.J., Loken C., Burns J.O., Hill J.M., White R.A., 1996, AJ, 112, 388 
\bibitem[\protect\citeauthoryear{Lubin, Mulchaey \& Postman} 
  {Lubin et al.}{2004}]{lubi04}  
  Lubin L.M., Mulchaey J.S., Postman M., 2004, ApJL, 601, L9  
\bibitem[\protect\citeauthoryear{Madgwick et al.}{2002}]{madg02}  
  Madgwick D.S.\ et al., 2002, MNRAS, 333, 133 
\bibitem[\protect\citeauthoryear{Mahdavi et al.}{1997}]{mahd97}  
  Mahdavi A., B{\"o}hringer H., Geller M.J., Ramella M., 1997, ApJ, 483, 68  
\bibitem[\protect\citeauthoryear{Mahdavi et al.}{2000}]{mahd00}  
  Mahdavi A., B{\"o}hringer H., Geller M.J., Ramella M., 2000, ApJ, 534, 114  
\bibitem[\protect\citeauthoryear{Mendes de Oliveira et al.}{2003}]{mend03}  
  Mendes de Oliveira C., Amram P., Plana H., Balkowski C., 2003, AJ, 126, 2635
\bibitem[\protect\citeauthoryear{Menon}{1992}]{meno92}  
  Menon T.K., 1992, MNRAS, 255, 41  
\bibitem[\protect\citeauthoryear{Merch{\'a}n \& Zandivarez}{2002}]{merc02} 
  Merch{\'a}n M., Zandivarez A., 2002, MNRAS, 335, 216  
\bibitem[\protect\citeauthoryear{Miles et al.}{2004}]{mile04}  
  Miles T.A., Raychaudhury S., Forbes D.A., Goudfrooij P., Ponman T.J., 
  Kozhurina-Platais V., 2004, MNRAS, 355, 785 
\bibitem[\protect\citeauthoryear{Mulchaey \& Zabludoff}{1998}]{mulc98} 
  Mulchaey J.S., Zabludoff A.I., 1998, ApJ, 496, 73  
\bibitem[\protect\citeauthoryear{Mulchaey et al.}{1996}]{mulc96} 
  Mulchaey J.S., Davis D.S., Mushotzky R.F., Burstein D., 1996, ApJ, 456, 80  
\bibitem[\protect\citeauthoryear{Mulchaey et al.}{2003}]{mulc03} 
  Mulchaey J.S., Davis D.S., Mushotzky R.F., Burstein D., 2003, ApJS, 145, 39 
\bibitem[\protect\citeauthoryear{Navarro, Frenk \& White} 
  {Navarro et al.}{1995}]{nava95} 
  Navarro J.F., Frenk C.S., White S.D.M., 1995, MNRAS, 275, 720  
\bibitem[\protect\citeauthoryear{Nicastro et al.}{2003}]{nica03}  
  Nicastro F.\ et al., 2003, Nature, 421, 719  
\bibitem[\protect\citeauthoryear{O'Sullivan, Ponman \& Collins} 
  {O'Sullivan et al.}{2003}]{osul03} 
  O'Sullivan E., Ponman T.J., Collins R.S., 2003, MNRAS, 340, 1375  
\bibitem[\protect\citeauthoryear{Osmond \& Ponman}{2004}]{osmo04} 
  Osmond J.~P.~F., Ponman T.~J., 2004, MNRAS, 350, 1511 
\bibitem[\protect\citeauthoryear{Pietsch et al.}{1997}]{piet97}  
  Pietsch W., Trinchieri G., Arp H., Sulentic J.W., 1997, A\&A, 322, 89  
\bibitem[\protect\citeauthoryear{Pierre et al.}{2004}]{pier04}  
  Pierre M.\ et al., 2004, JCAP, 9, 11  
\bibitem[\protect\citeauthoryear{Ponman et al.}{1996}]{ponm96}  
  Ponman T.J., Bourner P.D.J., Ebeling H., B\"{o}hringer H., 1996, MNRAS, 283, 
  690 
\bibitem[\protect\citeauthoryear{Popesso et al.}{2006}]{pope05}  
  Popesso P., Biviano A., B{\"o}hringer H., Romaniello M., 2006, A\&A,  
  in press (astro-ph/0606191) 
\bibitem[\protect\citeauthoryear{Ramella, Pisani \& Geller} 
  {Ramella et al.}{1997}]{rame97} 
  Ramella M., Pisani A., Geller, M.J., 1997, AJ, 113, 483  
\bibitem[\protect\citeauthoryear{Rasmussen \& Pedersen}{2001}]{rasm01} 
  Rasmussen J., Pedersen K., 2001, ApJ, 559, 892 
\bibitem[\protect\citeauthoryear{Rasmussen \& Ponman}{2004}]{rasm04} 
  Rasmussen J., Ponman T.J., 2004, MNRAS, 349, 722 
\bibitem[\protect\citeauthoryear{Read \& Ponman}{2003}]{read03} 
  Read A.M., Ponman T.J., 2003, A\&A, 409, 395  
\bibitem[\protect\citeauthoryear{S{\'e}rsic}{1974}]{sers74} 
  S{\'e}rsic J.L., 1974, Ap\&SS, 28, 365  
\bibitem[\protect\citeauthoryear{Sulentic, Pietsch \& Arp} 
  {Sulentic et al.}{1995}]{sule95} 
  Sulentic, J.W., Pietsch W., Arp H., 1995, A\&A, 298, 420  
\bibitem[\protect\citeauthoryear{Sutherland \& Dopita}{1993}]{suth93} 
  Sutherland R.S., Dopita M.A., 1993, ApJS, 88, 253  
\bibitem[\protect\citeauthoryear{Terashima et al.}{2002}]{tera02}  
  Terashima Y., Iyomoto N., Ho L.C., Ptak A.F., 2002, ApJS, 139, 1  
\bibitem[\protect\citeauthoryear{Trinchieri et al.}{2003}]{trin03} 
  Trinchieri G., Sulentic J., Breitschwerdt D., Pietsch W., 2003, A\&A, 401,  
  173 
\bibitem[\protect\citeauthoryear{Trinchieri et al.}{2005}]{trin05} 
  Trinchieri G., Sulentic J., Pietsch W., Breitschwerdt D., 2005, A\&A, 444,  
  697  
\bibitem[\protect\citeauthoryear{Tully}{1987}]{tull87}  
  Tully, R.B., 1987, ApJ, 321, 280  
\bibitem[\protect\citeauthoryear{van den Bergh}{1999}]{vand99}  
  van den Bergh S., 1999, A\&AR, 9, 273  
\bibitem[\protect\citeauthoryear{Verdes-Montenegro et al.}{1998}]{verd98} 
  Verdes-Montenegro L., Yun M.S., Perea J., del Olmo A., Ho P.T.P., 1998, ApJ, 
  497, 89 
\bibitem[\protect\citeauthoryear{White et al.}{1999}]{whit99} 
  White R.A., Bliton M., Bhavsar S.P., Bornmann P., Burns J.O., Ledlow M.J.,  
  Loken C., 1999, AJ, 118, 2014  
\bibitem[\protect\citeauthoryear{Zabludoff \& Mulchaey}{1998}]{zabl98} 
  Zabludoff A.I., Mulchaey J.S., 1998, ApJ, 496, 39  
\end{thebibliography}
\end{document}